\documentclass{aa} 
\usepackage{graphicx}
\usepackage{txfonts}
\newcommand{\etal}{et\,al.\ }
\newcommand{\logg}{\mbox{$\log g$}}
\newcommand{\Teff}{\mbox{$T_\mathrm{eff}$}}
\newcommand{\pgstar}{\object{HE\,1429$-$1209}}

\newcommand{\rxj}{\object{RX\,J2117.1$+$3412}}

\newcommand{\elf}{\object{PG1159$-$035}}

\begin{document}
   \title{Detection of non-radial g-mode pulsations in the newly discovered PG1159 star \pgstar}

   \author{T. Nagel \and K. Werner}

   \offprints{T. Nagel}

   \institute{Institut f\"ur Astronomie und Astrophysik, Abteilung
     Astronomie, Universit\"at T\"ubingen, Sand 1, 72076 T\"ubingen, Germany\\
     \email{nagel@astro.uni-tuebingen.de}
     }

   \date{Received xx; accepted xx}

   \abstract{We performed time-series photometry of the PG1159-type star
   \pgstar, which was recently discovered in the ESO SPY survey. We show that
   the star is a low-amplitude ($\approx$0.05\,mag) non-radial g-mode pulsator
   with a period of 919\,s. \pgstar\ is among the hottest known post-AGB
   stars (\Teff=160\,000\,K) and, together with the known pulsator \rxj, it
   defines empirically the blue edge of the GW~Vir instability strip in the HRD
   at high luminosities. 

   \keywords{stars: white dwarfs -- stars: individual: \pgstar -- stars:
   variables: GW~Vir -- stars: AGB and post-AGB } 
   }

   \maketitle
%

\begin{figure*}
  \centering \centering
  \includegraphics[width=12cm]{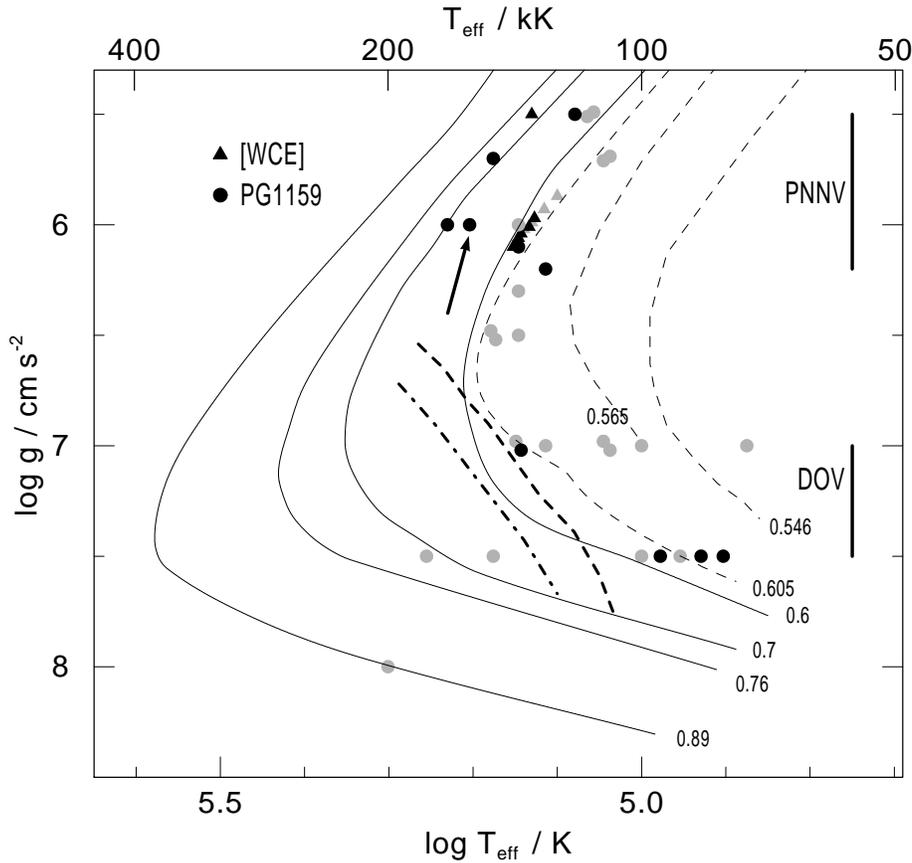}
  \caption{Location of the new pulsator (arrow) in the  \logg--$\log$\,\Teff\
    diagram among PG1159 stars (for parameters see Tab.\,\ref{pulsators}) and
[WCE] central stars. All of the known GW~Vir pulsators are plotted
with black symbols. According to their low and high surface gravity
(corresponding to high and low luminosity, respectively), they can be grouped
into two distinct classes of variable stars: planetary nebulae nuclei (PNNV) and
DO variables (DOV), as indicated by the vertical bars. Although the new pulsator
is within the PNNV region of the instability strip, no associated planetary nebula
is known. The thick dashed and dashed-dotted lines represent the theoretical
blue edge of the instability strip for l=1 and l=2 modes, respectively
(Gautschy 2004). Post-AGB evolutionary tracks are taken from Sch\"onberner (1983),
Bl\"ocker (1995) (dashed lines), and Wood \& Faulkner (1986) (solid lines) (labels: mass in
M$_\odot$).  } \label{fig_gteff}
\end{figure*}

\begin{figure*}
  \centering
  \includegraphics{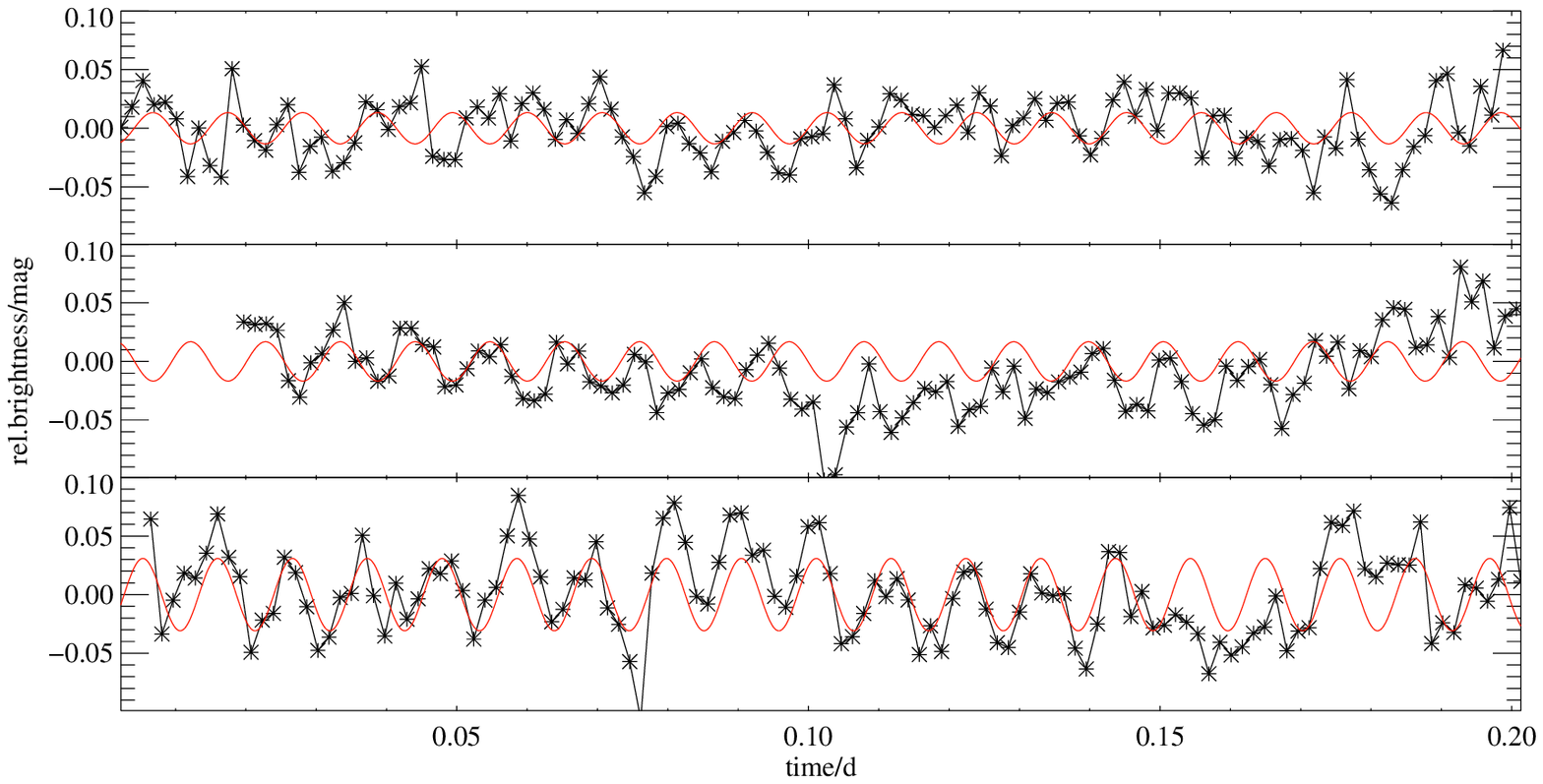}
  \caption{Lightcurves (rebinned by factor of 2) of \pgstar\ during three
  consecutive nights (from top to bottom) and a sine fit with the detected
    period of 919\,s. The time is measured in fractions of a day, beginning with
    the start of observations.}\label{binned_lightcurve}
\end{figure*}

\section{Introduction}

PG1159 stars are hot hydrogen-deficient (pre-) white dwarfs ($T_{\rm
eff}$=75\,000--200\,000\,K, $\log g$=5.5--8 [cm/s$^2$]; Werner 2001). They are
probably the outcome of a late helium-shell flash, a phenomenon that drives the
currently observed fast evolutionary rates of three well-known objects (FG~Sge,
Sakurai's object, V605 Aql). Flash-induced envelope mixing produces a
H-deficient stellar surface. The photospheric composition then essentially
reflects that of the region between the H- and He-burning shells in the
precursor AGB star. The He-shell flash forces the star back to the AGB. The
subsequent, second post-AGB evolution explains the existence of Wolf-Rayet
central stars of planetary nebulae and their successors, the PG1159 stars.

Ten out of the 32 currently known PG1159 stars (including the object studied
here) are non-radial g-mode pulsators (Tab.\,\ref{pulsators}).  They are of
considerable interest for asteroseismology studies, because one can reveal the
interior stellar structure and thus obtain important hints as to the
evolutionary history of these peculiar stars. The pulsators form the group of
GW~Vir variables, named after the prototype GW~Vir (=\elf). Sometimes this group
is divided into two subgroups, termed PNNV (Planetary Nebula Nuclei Variables)
and DOV (DO Variables; DOs are hot white dwarfs with He-rich atmospheres). The
six PNNV are low-gravity, hence high-luminosity stars, whereas the four DOV are
of high gravity (Fig.\,\ref{fig_gteff}). This subdivision is somewhat artificial,
because the physical pulsation driving mechanism is identical for all GW~Vir
variables. In addition, two of the objects located within the PNNV region of the
\logg--$\log$\,\Teff\ diagram  have no known associated planetary nebula
(HS2324+3944 and \pgstar). We also mention that in Fig.\,\ref{fig_gteff} we have
plotted the location of [WCE] (early-type C-rich Wolf-Rayet) central stars, which are thought to be the
progenitors of PG1159 stars. Among them, we marked six pulsators (NGC\,6905,
Sand\,3, NGC\,2867, NGC\,5189, NGC\,1501, NGC\,2371), which in addition to the
six PG1159 stars mentioned above are also termed PNN variables (Ciardullo \&
Bond 1996; atmospheric parameters were taken from Hamann 1997 and Herald \&
Bianchi 2004).  It seems obvious that these Wolf-Rayet central stars also
belong to the GW~Vir variables in a sense that their pulsation driving mechanism
is identical.

The GW~Vir variables define a new instability strip in the HRD. The driving
mechanism has been discussed in the past with considerable controversy. The
basic ingredient, the $\kappa$-mechanism associated with the partial ionization
zone of K-shell electrons of C and O in stellar model envelope just beneath the
photosphere, has been identified in  the pioneering work by Starrfield \etal
(1983). However, these models required a very He-deficient composition, in
contrast to the observed surface chemistry. Therefore a composition gradient had
to be assumed, which is difficult to understand considering on-going mass-loss
from these objects, what prevents gravitational settling of heavy elements. In
contrast, later work by Saio (1996), Gautschy (1997), and Quirion \etal (2004)
showed that pulsations can easily be driven in chemically homogeneous envelopes
with abundances according to spectroscopic analyses. In particular, Quirion
\etal (2004) can successfully explain pulsation properties of individual PG1159
stars. But still, these results are at odds with conclusions from Bradley \&
Dziembowski (1996) and Cox (2003), which both require different envelope and
atmospheric compositions.

In view of this controversy and the relatively small number of GW~Vir variables
spread out over a wide range in the  \logg--$\log$\,\Teff\ diagram, the
discovery of a new member of this class is interesting for the question of
pulsation driving of these variables and asteroseismology investigation of
the PG1159 group as a whole. 

Recently, the discovery of a new PG1159 star, \pgstar, within the
SPY (Supernova~Ia  Progenitor Survey, Napiwotzki \etal 2003) has been announced
(Werner \etal 2004a). It is among the hottest objects within this group and has
a low surface gravity (\Teff=160\,000\,K, \logg=6), hence it is located within
the GW~Vir instability strip, right among the PNN variables. This fact along
with a high carbon abundance (C=54\%, He=38\%, O=6\%, Ne=2\%, by mass),  which
is a prerequisite for pulsation driving within the instability strip (Quirion
\etal 2004), strongly suggested that \pgstar\ might be a new GW~Vir pulsator.

\section{Photometry of \pgstar}

Photometric observations of \pgstar\ ($m_{\rm V}=16.1$) were performed during three consecutive
nights (Tab.\,\ref{tab_obs}) using our institute's 0.8\,m f/8 telescope with an
SBIG ST-7E CCD camera. To achieve good time resolution  we chose clear filter
exposures with a binning of 2x2 pixels to reduce readout time.  The exposure
time was T$_{\rm exp}$=60\,s and the readout time 8\,s, resulting in a cycle
time of T$_{\rm cycl}$=68\,s. Data were obtained over almost five hours per
night.   The observing conditions have been good during all three nights,
considering that the telescope is located in the city of T\"ubingen.

\begin{table*}
\centering
\caption{Parameters of all known pulsating PG1159 stars (corresponding to filled dots in
  Fig.\,\ref{fig_gteff}), including the new pulsator \pgstar. See text for references.}
\begin{tabular}{lrcccc}
      \hline
      \hline
      \noalign{\smallskip}
     & \Teff     &  \logg     & $M $          & Unstable Periods & Observed Periods \\
Name & (1000\,K) & (cm/s$^2$) & (M$_\odot$) & (s)              & (s) \\
      \noalign{\smallskip}
      \hline
      \noalign{\smallskip}
PG\,0122+200 &  80 & 7.5 & 0.50 &  202--822  &  336--612 \\
PG\,1159-035 & 140 & 7.0 & 0.53 &  483--704  &  430--840 \\
PG\,1707+427 &  85 & 7.5 & 0.52 &  200--778  &  336--942 \\
\rxj         & 170 & 6.0 & 0.70 & 1289--2332 &  694--1530 \\
PG\,2131+066 &  95 & 7.5 & 0.55 &  191--643  &  339--598 \\
HS\,2324+3944& 130 & 6.2 & 0.55 & 1355--3013 & 2005--2569 \\
K1-16        & 140 & 6.4 & 0.50 &  857--2192 & 1500--1700 \\
NGC\,246     & 150 & 5.7 & 0.70 & 1160--1466 & 1460--1840 \\
Longmore 4   & 120 & 5.5 & 0.65 &            &  831--2325 \\
\pgstar      & 160 & 6.0 & 0.68 &            & 919 \\
 \noalign{\smallskip}
      \hline
\end{tabular}\label{pulsators}
\end{table*}

Data reduction was done with our IDL software TRIPP (Time Resolved Imaging
Photometry Package, Schuh \etal 2003), performing aperture photometry. All
images are background and flatfield corrected. The relative flux of the
object is calculated with respect to one or more comparison stars.  To
minimize flux errors, different aperture radii were tested in respect of
the comparison star flux variances.  This procedure allows the detection
of very small brightness variations.  The resulting lightcurve is displayed in
Fig.\,\ref{binned_lightcurve}.

To analyse the combined lightcurve of the three nights, we used CAFE (Common
Astronomical Fit Environment, G\"ohler, priv. comm.), a sample of routines
written in IDL.

\begin{figure}
  \centering
  \includegraphics[width=9cm]{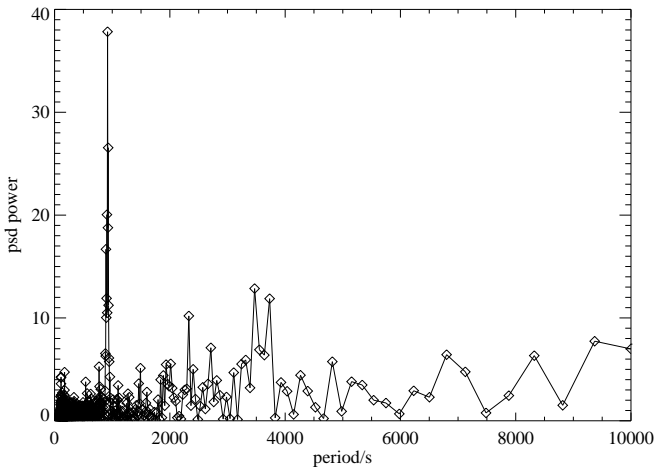}
\caption{Lomb-Scargle periodogram of the combined lightcurve of \pgstar. The
  power spectral density (psd) is a measure for the probability of a period
  being present in the lightcurve. The confidence limit of 99\,\% is
  at psd=11.8.
}\label{lomb} 
\end{figure}

\begin{figure}
  \centering
  \includegraphics[width=9cm]{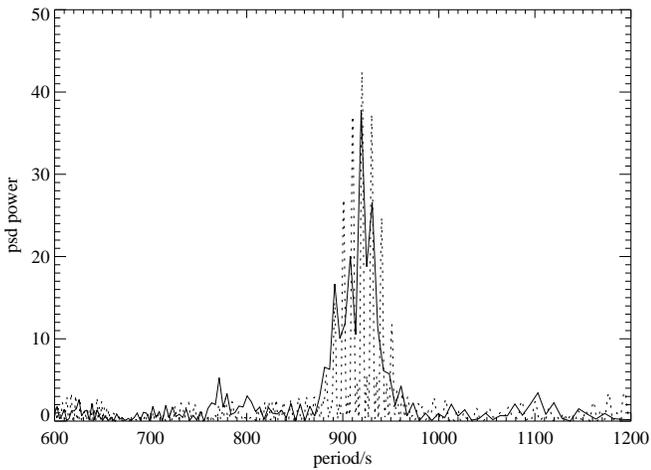}
  \caption{Period spectrum (solid line) and spectral window (dotted line)
  of the combined lightcurve of \pgstar.}\label{window} 
\end{figure}

We calculated a Lomb-Scargle periodogram (Scargle 1982) and simulated a
lightcurve with the same time sampling and noise characteristics as the
observations and with the suspected period in order to get the spectral window.
In Fig.\,\ref{lomb} the Lomb-Scargle periodogram of the combined lightcurve of
the three nights is presented. A peak which exceeds a 99\% false-alarm
probability (which was estimated according a prescription of Scargle et
al. 1982) of 11.8 by a factor of three indicates a pulsation period of
919\,s for \pgstar. To check the period for
artefacts, we simulated a lightcurve with the same statistical behaviour
like the observed lightcurve. Its spectral window is shown in
Fig.\,\ref{window} as well as the periodogram of the lightcurves.  In
Fig.\,\ref{binned_lightcurve} the binned lightcurves of \pgstar\ and a sine
fit with the detected period of 919\,s are compared. The observed
variability has a mean  amplitude of about 0.05\,mag, with significant
variation from night to night, indicating the possible presence of
additional beat periods.

\begin{table}
\centering
\caption{Observation log}
\begin{tabular}{rcccc}
      \hline
      \hline
      \noalign{\smallskip}
 Date     & UT    & T$_{\rm exp}$/s   & T$_{\rm cycl}$/s & Duration/s\\
      \noalign{\smallskip}
      \hline
      \noalign{\smallskip}
22.05.2004 & 20:17 & 60           & 68          & 17799\\
23.05.2004 & 20:20 & 60           & 68          & 17485\\
24.05.2004 & 20:18 & 60           & 68          & 17312\\
 \noalign{\smallskip}
      \hline
\end{tabular}\label{tab_obs}
\end{table}

\section{Discussion}

We have discovered a 919\,s period in the lightcurve of the recently identified
PG1159 star \pgstar\ (\Teff=160\,000\,K, \logg=6). Its location among other
pulsators in the GW~Vir instability strip suggests that the star is a non-radial
g-mode pulsator. Together with \rxj\ (\Teff=170\,000\,K, \logg=6) this new
pulsator defines empirically the blue edge of the GW\,Vir instability strip at
low gravities, i.e., at high luminosities in the HRD. This is in complete
agreement with recent pulsation modeling (Quirion \etal 2004, Gautschy 2004). In
Fig.\,\ref{fig_gteff} we have plotted the location of the blue edge, recently
derived by Gautschy (2004). The high amount of carbon found in \pgstar\ (Werner
\etal 2004a) confirms the conclusion of Quirion \etal (2004) that the
co-existence of non-pulsating stars in the instability strip is due to a low
carbon abundance relative to helium in the non-pulsators, i.e., helium-poisoning
of pulsations.

The atmospheric parameters of \pgstar\ are, within error limits, identical to
those of \rxj\ and thus, we can compare the detected 919\,s period with the
pulsation periods found in that star and with the respective model calculations
presented by Quirion \etal (2004). In Table\,\ref{pulsators} we list stellar
parameters as well as observed and predicted pulsation periods for all ten known
pulsating PG1159 stars. The table was essentially taken from Quirion \etal (2004, see
references therein) and it is augmented by the central star Longmore~4 and the
new pulsator. Parameters and observed pulsation periods of Longmore~4 are taken
from Werner \etal (2004b) and Bond \& Meakes (1990), respectively. But note that
we speculated in the referenced work that \Teff\ is probably
underestimated. Obviously, the detected 919\,s period in \pgstar\ is within
the intervals of observed and predicted periods of \rxj, supporting our
discovery of g-mode pulsations.

Future observations of \pgstar\ during longer, uninterrupted time intervals will
almost certainly reveal other, weaker pulsation periods which will allow a
detailed study of the interior structure of this star with asteroseismology
tools.

\begin{acknowledgements}
We thank Sonja Schuh (G\"ottingen) and Eckart G\"ohler (T\"ubingen) for helpful
discussions, and Alfred Gautschy (Basel) for allowing us to present his
calculated blue edges of the GW~Vir strip in advance of publication.
\end{acknowledgements}

\end{document}